# Moving towards high-power thin-disk lasers in the 2-µm wavelength range

SERGEI TOMILOV*, MARTIN HOFFMANN, YICHENG WANG AND CLARA J. SARACENO

*Sergei.Tomilov@ruhr-uni-bochum.de

Abstract:

Thin-disk lasers (TDLs) have made spectacular progress in the last decades both in continuous-wave and ultrafast operation. Nowadays, single thin-disk oscillators with > 16 kW of continuous-wave (CW)-power have been demonstrated and ultrafast amplifiers have largely surpassed the kilowatt milestone with pulse energies in the multi-100 mJ range. This amazing development has been demonstrated in the 1-µm wavelength range, using Yb-doped materials and supported by industrially available components. Motivated by both strong scientific and industrial applications, interest in expanding this performance to longer wavelength regions continues to increase. In particular, TDLs emitting directly in the short-wave mid-infrared (SW-MIR) region (2-3 µm) are especially sought after, and although many early studies have been reported, most remained in the proof-of-principle stage and the potential for multi-100-W operation remained undemonstrated. Here, we report on our recent results of a single fundamental-mode CW Ho:YAG thin-disk oscillator with >100 W of power, surpassing previous single-mode TDLs by a factor of >4, and marking a first milestone in the development of high-power SW-MIR TDLs. In optimized conditions, our laser system emitting at $\approx$ 2.1 µm reaches an output power of 112 W with 54.6-% optical-to-optical efficiency and an $M^2$ = 1.1. This system is ideally suited for future direct modelocking at the 100 W level, as well as for ultrafast amplification. We start the discussion with a review of the state-of-the-art of TDLs emitting directly in the vicinity of 2 µm, and then discuss difficulties and possible routes both towards ultrafast operation and next possible steps for power scaling.

## 1. Introduction and state-of-the-art

High-power lasers in the SW-MIR wavelength region (2-3 µm) are of great interest for a large variety of applications in both science and industry. One example in industry is their potential for laser material processing of exotic materials, for example laser welding of polymers [1]. In scientific research, finding paths to increase the average power of ultrafast laser systems emitting in this wavelength region has seen particularly strong interest, since these sources have great potential as efficient drivers for secondary sources in the XUV, mid-IR and THz [2–6].

So far, accessing the SW-MIR wavelength region with watt-level average power and above was mostly achieved with complex parametric amplifiers, pumped by better established high-power systems in the 1 µm wavelength region [3,7]. However, this results in complex and inefficient systems. In contrast, laser systems based on materials offering laser transitions emitting directly at such wavelengths offer an extremely elegant and simple solution and are currently the topic of extensive investigation in the laser community. In this respect, $Tm^{3+}$ or $Ho^{3+}$ ions offer great advantages in comparison to other SW-MIR active media. Pioneering studies on $Tm^{3+}$ and $Ho^{3+}$ gain media by Johnson et al. go back to 1962 [8]. Both ions operate in the quasi-3-level scheme with laser transitions in the 2-µm wavelength range ($^3F_4 \rightarrow {}^3H_6$ in $Tm^{3+}$, $^5I_7 \rightarrow {}^5I_8$ in $Ho^{3+}$) [9,10]. $Tm^{3+}$ ions can be directly pumped by commercially available high-power AlGaAs based diodes at ~790 nm and emit at ~2.0 µm with high-efficiency due to the cross-relaxation process between $Tm^{3+}$-$Tm^{3+}$ ions ($^3H_4 + {}^3H_6 \rightarrow 2 \times {}^3F_4$), resulting in ~80 % of theoretical slope efficiency, thus making it very well suited for high-power applications. On the other hand, $Ho^{3+}$ ions require in-band pumping at ~1.9 µm but emit at ~2.1 µm, featuring longer wavelength and higher theoretical stokes efficiency of nearly 91 %, again showing great potential for high-power operation. Since commercially available diode lasers at 1.9 µm are still rare and very expensive [11], the most common pumping scheme for pumping $Ho^{3+}$-doped materials is still $Tm^{3+}$-fiber lasers. Alternatively, ($Tm^{3+}$, $Ho^{3+}$) co-doped media offer another alternative for diode-pumping, where diode-pumped $Tm^{3+}$ ions transmit their energy to the emitting $Ho^{3+}$ ions through the lattice [12]. Often, these materials exhibit broader emission bandwidths than their single dopant counterparts, making it also attractive for ultrafast operation [13].

Besides $Tm^{3+}$ and $Ho^{3+}$, transition-metal doped II-VI semiconductor materials, such as $Cr^{2+}$:ZnS and ZnSe, emitting at 2-3.5 µm, have also attracted attention for SW-MIR operation, and are known as 'the Ti-Sapphire of the mid-IR'. Great progress has been achieved with Cr:ZnS/ZnSe [10,14–16], with CW powers up to 140 W in bulk amplifier operation, however requiring a sophisticated beam rotation scheme [17] and broadband modelocked bulk oscillators with watt-level operation, potentially supporting sub-20 fs [15]. With bulk amplification systems, 7.4-W average power was also demonstrated at a repetition rate of 81 MHz, with pulses as short as 40 fs centered at 2.4-µm wavelength [15].

With this large potential in mind, many power-scalable laser geometries are currently being investigated using these active ions, including thin-disk, fiber and slab lasers. In particular, $Tm^{3+}$ and $Ho^{3+}$ are currently heavily explored, since Cr-doped transition metals have a more limited commercial availability in high quality, as well as well-known difficulties for extreme power scaling due to very large thermal lensing [18,19]. We therefore focus here our attention on most recent progress achieved in these geometries using $Tm^{3+}$ and $Ho^{3+}$-doped materials.

Fiber lasers have made large progress using $Tm^{3+}$-glass. In 2009, a 415 W single mode CW Tm-fiber laser (in-band pumped by Er-fiber) at 1940 nm has been demonstrated with a slope efficiency of 60 % and optical efficiency of 25 % [20]. In the same year, Moulton et al. demonstrated a diode-pumped $Tm^{3+}$-fiber laser with an output power up to 885 W at ~2040 nm [21] with an optical-to-optical efficiency of 51 %. These diode-pumped Tm-fiber lasers later developed into kW-class systems. At the CLEO conference in 2013, a single-mode $Ho^{3+}$-doped fiber laser cladding-pumped by $Tm^{3+}$-fiber laser has been demonstrated, delivering an average output power up to 407 W at 2.12 µm [22]. Further advances have also been recently made in ultrafast operation, reaching the kW level [23]. With slab lasers, an output power of 148 W was achieved with Tm:YLF in 2009 and 65 W with Ho:YAG in 2017 [24,25], and further scaling remains to be explored.

Compared to fiber lasers, TDLs have made comparably little progress in this wavelength region, making this an area of enormous potential. In fact, as shown by the spectacular recent development of 1 µm systems [26], TDLs offer several unique advantages inherent to the disk geometry: on the one hand, significantly higher power levels with excellent beam quality are achievable from a single-gain element without yet requiring coherent combination [27], on the other hand, for ultrafast operation, extremely high-energy single stage amplifiers based on regenerative and multi-pass amplification are achievable [28,29], as well as high-power modelocked lasers [30]. In the 1-µm wavelength region, up to 206 mJ and 1.4 kW have been demonstrated using regenerative amplifiers [29], and 4.7 mJ and 1.4 kW using multi-pass amplifiers [28]. Moreover, single-box modelocked oscillators have achieved up to 350-W average power [30] and 80-µJ pulse energies [31].

Extending this unique performance to the 2-µm range is thus of high interest for the laser community. However, so far only very few results have conclusively shown this potential as illustrated in the overview graph in Fig. 1. The first 2-µm TDL was reported at the CLEO conference in 1998, only 4 years after the invention of TDLs [32]. In this experiment, a 500-µm 10-at.% Tm:YAG crystal was diode pumped at 785 nm with 8 passes and cooled by a Peltier-cooled heat sink achieving 2 W of CW power, and 18.3 W of peak power in quasi-CW regime [33]. This was improved to 6 W one year later, with slightly lower doping (6-at.% Tm:YAG) and thicker disks (650 µm) [34]. In 2006, the first $Ho^{3+}$-based TDL was demonstrated with 400 and 500-µm thick disks, ~2-at.% Ho:YAG, reaching average output power of 9.4 W with an optical-to-optical efficiency of 36 % [35]. Multi-pass pumping scheme with 24 passes on the disks was introduced to 2 µm TDLs in this work, using a diode-pumped Tm:YLF laser as pumping laser. Additionally to this early work on YAG based disks, 2-µm TDLs were also explored with other host materials including Tm/Ho:KYW [36,37], Tm:LLF [38], $Tm:Lu_2O_3$ [39], Tm:KLuW [40] as well as Cr:ZnSe [18,11,41]. This resulted in power levels of 21 W with Tm:LLF [38] and 22 W with InP diode-pumped Ho:YAG [11]. More recently, multimode CW operation up to 24 W ($M^2$~11.7) using Tm:YAG and 50 W ($M^2$~5.3) using Ho:YAG was demonstrated [42], and a first Ho:YAG Kerr-lens modelocked TDL with an average power up 25 W and 220 fs pulses was achieved [43]. These results were using most recent advances in thin-disk components, including a 72-pass pump head, and a disk attached to a diamond heat sink (TRUMPF laser GmbH).

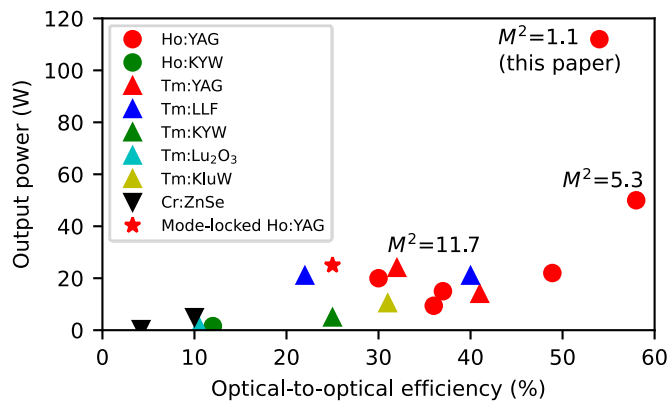

Fig. 1. Overview of 2 μm CW and modelocked TDLs (available $M^2$ data included) [11,18,35–45]

Here, we conclusively show that the technology has significantly matured, and is reaching a promising turning point. We demonstrate a $Tm^{3+}$-fiber in-band pumped Ho:YAG thin-disk oscillator operating in fundamental-mode CW regime emitting a record high-power of 112 W with high optical-to-optical efficiency of 54.6 % and $M^2$~1.1 with dual wavelength emission at 2090 nm and 2096 nm. To the best of our knowledge, this is the highest CW power achieved with a TDL in this wavelength region, surpassing the 100-W milestone. We discuss the next step towards ultrafast modelocked operation and further power-scaling, based on our investigation of this system's performance.

## 2. 100-W class fundamental-mode thin-disk laser

The experimental setup of the 2-μm high-power TDL is shown in Fig. 2. A 190-μm thick Ho:YAG disk with 2-m radius of curvature (RoC) and 10-mm diameter is glued to a water-cooled diamond heatsink at a temperature of 18 °C. During the experiments, two disks with 1-at.% and 2-at.% $Ho^{3+}$ doping concentrations were tested. A multi-pass pumping head featuring 72 pump passes through the disk was used to achieve high pump absorption (TRUMPF). The pump was a Tm-fiber laser (IPG Photonics) with an emission wavelength of 1908 nm and maximum output power of 210 W. Due to the absence of commercially available coupling fibers for this wavelength range, we implemented free-space coupling using two lenses with focal lengths of 229 mm and 26 mm to image the flat-top pump beam to the disk. With a pump beam radius of 1.2 mm, a maximum pump power intensity of 4.5 kW/cm² was applied to the disk. At this pump intensity, the surface disk temperature was slightly above 100 °C (measured with a thermal camera), comparable to values that are typically observed in Yb:YAG disks at these pump intensities.

The V-shaped laser cavity is formed by a curved high-reflective spherical mirror with 1-m RoC, the thin-disk active medium and a flat output coupler (OC). The total cavity length is 1.87 m. To ensure single-mode operation of the laser, the resonator was designed to provide a mode size on the disk approximately 1.5 times bigger than the pump spot. This ratio was experimentally obtained, by scanning the curved mirror to find the ideal balance between optical efficiency and beam quality. In this resonator, we use the pump spot as a soft aperture which results in lower gain for the higher order modes in comparison to the TEM$_{00}$ mode. Besides that, no other mode-cleaning methods were used.

Most of the laser output was directed to a power meter using an OC with a high reflectivity of 99.6 %. The remaining transmitted beam (0.4 %) was used for different diagnostics (spectrum analyzer and $M^2$-measurement).

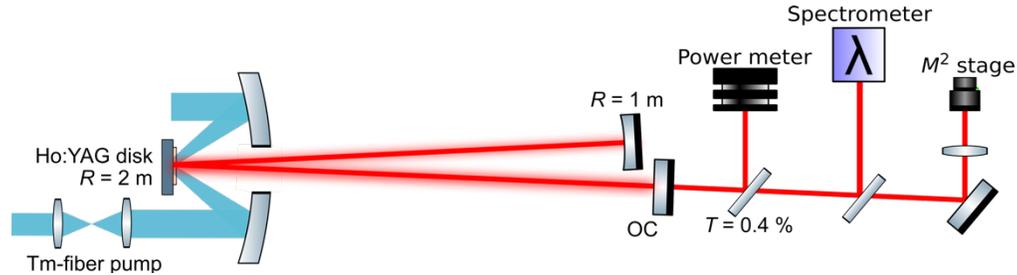

Fig. 2. Laser cavity and characterization setup

In Fig. 3, we show the different measurements realized using the two doping concentrations available at the time of the experiment and varying OC coefficients. Since the two 1-at.% and 2-at.% disks have the same thicknesses, one might expect better performance from the 2-at.% doped disk due to the higher pump absorption and higher available gain. Nevertheless we seek here to have better understanding of the influence of doping concentration in the disk geometry, since Ho:YAG is well-known to exhibit deleterious non-radiative transitions, that can degrade laser performance at higher-doping concentrations. It is worth noting that, as we point out in our discussion section later, the tolerable doping concentration limits in the thin-disk geometry have so far not been thoroughly investigated.

The power slopes of our 1-at.% doped disk were performed using three different OCs with transmission coefficients of 1, 2 and 3 % (Fig. 3a). In this case, the low doping concentration of the active medium led to a rather poor absorption of the pump radiation within the disk (82 % against 97 % in case of the 2-at.% disk), which resulted in high back-reflected power. We therefore typically limited our pump power to < 100 W to avoid damage of our pump laser. It is worth noting that this practical issue can be circumvented with other thin-disk pumping geometries without a retro-reflective mirror [46], in case lower doping concentrations are desired, and absorption efficiency can be compromised. The best slope efficiency of 48.8 % was achieved with a 1-% OC, with a maximum output power of 50 W. Experiments with other OCs resulted in worse slope efficiencies. Up to this pumping level, no saturation effects were observed even at the highest intracavity power of 5 kW.

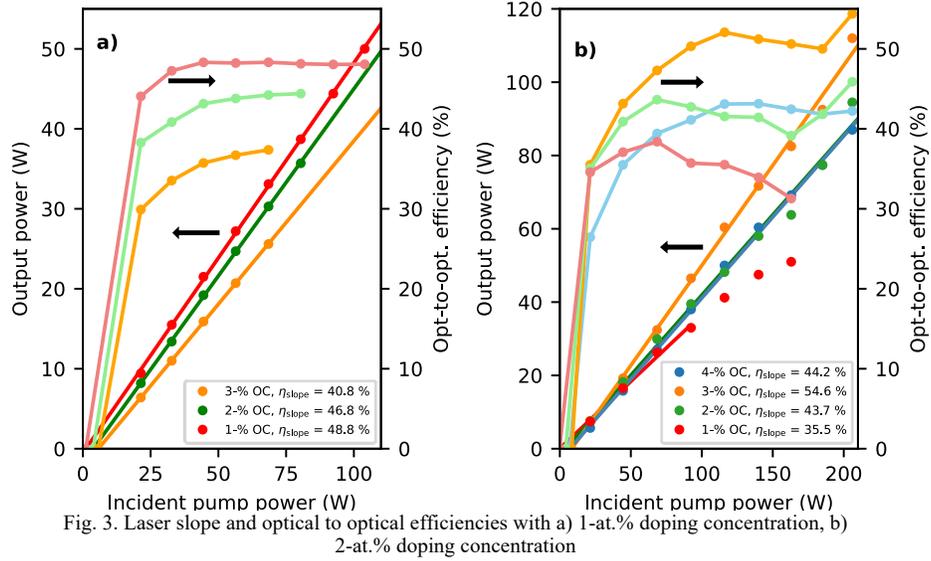

Fig. 3. Laser slope and optical to optical efficiencies with a) 1-at.% doping concentration, b) 2-at.% doping concentration

We repeated the same experiments for another disk with a higher doping concentration of 2 at.%, obtaining significantly better results (Fig. 3b). The best slope efficiency value of 54.6 % was achieved using an OC with a transmission coefficient of 3 %. In this configuration, the absorption of the disk was sufficient to use the full pump power.

The output power increases linearly with the pump power showing no over-saturation in the 2-%, 3-% and 4-% OC cases. The small change in optical-to-optical efficiency obtained at highest powers is explained by realignment of the cavity at maximum power. In the case of 1-% OC, saturation is most likely due to thermal effects of different cavity elements due to the extremely high intracavity power of more than 5 kW. The best performance was reached with a 3-% output coupling coefficient at 2 at.% of doping. At the full pump power of 210 W, we reach a maximum output power of 112 W, corresponding to an intra-cavity power of 3.73 kW. At this power, moderate heating of intracavity mirrors optical coatings was observed, as shown in Fig. 4, where a high-reflector is shown with an intensity of 1.5 MW/cm$^2$.

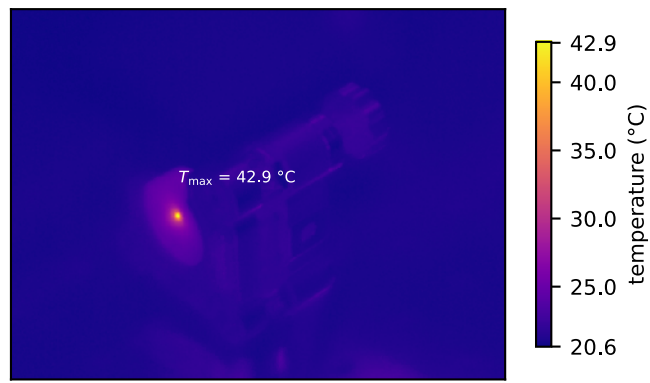

Fig. 4. Thermal image of the curved end mirror with beam intensity of 1.5 MW/cm$^2$.

The laser operated stably without significant thermal degradation on a time scale of at least one hour.

We measured the optical spectrum at the maximum power with different OCs and both disk concentrations in order to observe the influence of intracavity losses on the laser gain profile of Ho:YAG. As shown in Fig. 5, the laser emits two peaks at 2090 nm and 2096 nm. This double-wavelength operation was observed with all tested OC's within the full pump laser operation range and is a typical signature of spatial hole burning in the active medium [47]. The peaks balance shift is a consequence of the structured gain profile of Ho:YAG [48], with the highest peak being shifted toward shorter wavelength when losses increase. Fig. 5 shows the normalized gain profile of the active medium with two different excited ion ratios $β$, overlapped with the measured spectra. As expected, our measurements show that a higher OC transmission coefficient, as well as a lower active medium doping concentration (i.e. a larger inversion ratio $β$) shifts the balance between CW peaks towards the shorter wavelength peak at 2090 nm.

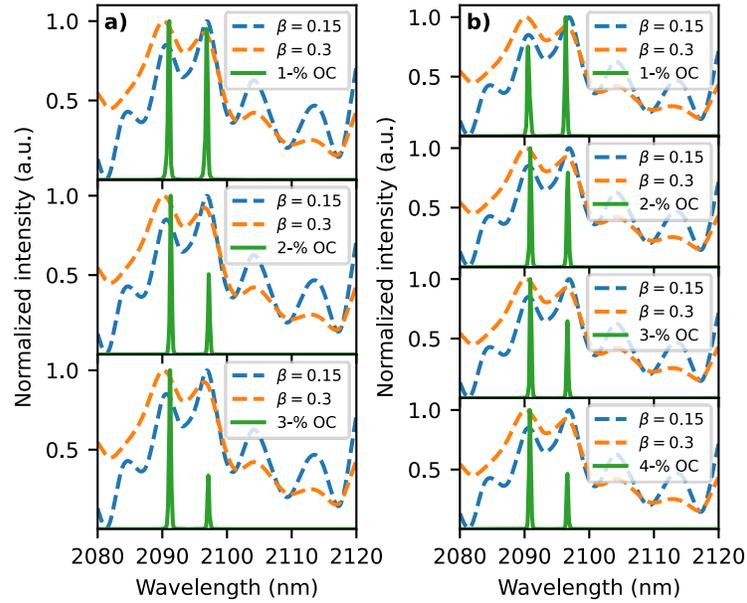

Fig. 5. Laser spectrum at different intracavity loss levels for a) 1-at.% doping concentration, b) 2-at.% doping concentration

Additionally, we calculated the intracavity losses with the approximated slope efficiency values for different OCs using the Caird analysis [49]. For this purpose, the $1/η$ dependence on the inverse of the OC transmission coefficient $1/T$ was calculated as shown in Fig. 6.

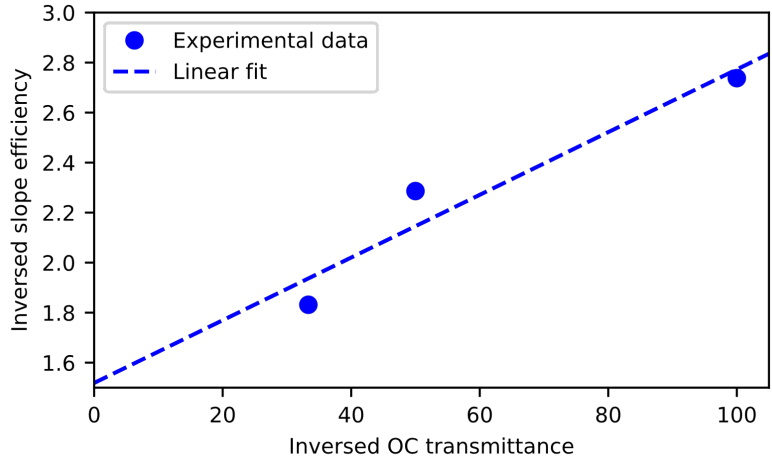

Fig. 6. Dependence of inverted slope efficiency of the laser with 2-at.% doped disk as a function of inverted OC transmittance

According to [49], the net losses in the cavity can be found using the following expression:

$$\frac{1}{\eta} = \frac{1}{\eta_0} + \frac{L}{\eta_0} \cdot \frac{1}{T},$$

where $\eta_0$ is the theoretical maximum slope efficiency and $L$ the intracavity loss coefficient. The calculated value for the internal losses was $L = 0.8\ \%$, indicating the excellent quality of the active medium and the mirrors, and is very promising for further modelocking experiments.

In order to characterize the beam quality of the laser, we measured the beam profile with a scanning-slit camera and performed an $M^2$ measurement of the laser with the 2-at.% doped disk at the maximum power with a 3-% OC. No changes of the beam and the output power were observed during the full power range. The measurements were done by carefully introducing an additional OC and fused-silica wedges into the beam path so that no additional thermal effects influence the measurements. A 150-mm lens was placed before the slit-scanner camera to form a steep caustic.

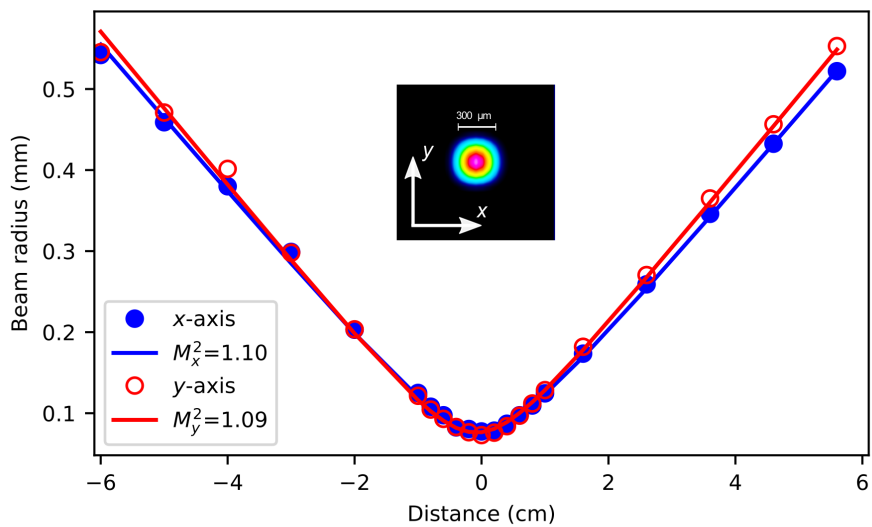

Fig. 7. Beam caustic (inset) and $M^2$ measurement results

We measured an excellent beam quality with $M^2$ values of 1.10 along the *x*-axis and 1.09 along *y*-axis as shown in the Fig. 7. These values are suitable for future mode-locking experiments. Such beam quality at full output power indicates a robust cavity design as well as excellent thermal properties of the intracavity optical elements, which are typically problematic in this wavelength range.

$M^2$ measurements were also carried out for the laser with the 1-at.% doped disk at 50 W power using the same procedure and resulted in the values of 1.43 along *x*-axis and 1.22 along *y*-axis. Deterioration of beam quality occurred only when the intracavity power exceeded 5 kW and remains Gaussian-shaped at lower pump powers, however we only performed $M^2$ measurements only at maximum intracavity intensity values, due to the poor performance of this disk for future experiments.

## 3. Discussion and outlook

3.1 Further power scaling

The laser demonstrated in the previous section is to the best of our knowledge the most powerful 2-µm TDL demonstrated so far. Nevertheless, further progress is required to reach the state-of-art of 1-µm TDLs, as well as that of $Tm^{3+}$ and $Ho^{3+}$ doped fibers in terms of the maximum output power, and we aim to discuss here short- and long-term perspectives to further power scaling.

The most straight-forward path to higher power levels is to make use of the power-scaling concept of TDL, i.e. increasing the pumping power and pumping area by the same factor. In this respect, the lack of cost-effective commercial $Tm^{3+}$-fiber lasers with higher output powers remains one key limiting factor. kW-class pumps exist but are extremely costly compared to diode pumps and are rare on the market. While we certainly believe that this is a matter of time given the current high interest in developing 2-µm sources, we focus our discussion here on how to use existing pumping sources and technical capabilities, to discuss near-future improvements.

One straight-forward step is to optimize the disk thickness and doping concentration to obtain the highest possible single-pass gain from one gain unit. It is worth noting that in most previous studies, multi-pass pumping with only 24 passes was used, and thus thicker disks were necessary, which leads to significantly higher operation temperatures and thermal effects. Nowadays, state-of-the-art pumping heads with 72 pump passes can be used, contacting of disks on diamond with excellent quality is possible, therefore much thinner disks with much better heat removal can also operate efficiently, which significantly reduces detrimental heating effects. A thorough investigation of temperature rise and laser efficiency in a much larger parameter space than presented above is needed. Such a study can be expected in the future. In particular, the limits in terms of doping concentration in the disk geometry are of critical importance, as we highlight below.

One important point to note is that our result highlights that current 2-µm optical components for CW operation are starting to be at comparable performance level as their near-infrared counterparts. At highest output power, our Ho:YAG laser operates with 3.7 kW of intracavity power without significant thermal degradation, as expected from the low intracavity losses evaluated in our experiments. At intracavity powers approached 5 kW, thermal effects were observed, but future improvements in coating technology for this wavelength region are expected to facilitate operation, even at these extreme levels, in the near future. Nevertheless, laser long-term stability is improved when operating at lower intracavity powers, for example by increasing the number of gain passes on the disk. In view of future ultrafast operation, lower intracavity power is also desired to reduce nonlinearities and thermal load on more critical components, as emphasized below and discussed in detail in [50]. This technique has been extensively used for modelocking in the 1-µm region [51–53], however often at the expense of a more complex layout and typically a higher sensitivity to thermal lensing effects. Next power-scaling steps at 2 µm will certainly be a combination of an optimized single-stage gain element, and multiple passes to increase the ratio between output and intracavity power. We emphasize here additionally, that this will be of critical importance to achieve power scaling in ultrafast operation, where more lossy components (dispersive mirrors and saturable absorbers) will be required intracavity, making a higher gain and lower thermal effects much more critical than in the CW case. In the following section, we discuss the first step in optimizing the single-gain per pass via an increase of doping concentration.

### 3.2 Higher doping concentrations

Considering the results demonstrated in the previous section, it is of interest to discuss in more detail the influence of a higher doping concentration in Ho:YAG lasers in the thin-disk geometry. At a given thickness, highest possible doping levels are beneficial because it allows us to increase the pump absorption and gain per pass, thus allowing us to achieve better efficiency, extract more power at lower intracavity power or simply tolerate higher intracavity loss, all of which are beneficial for our final target of achieving 100-W class ultrafast operation. A balance naturally needs to be found with the fact that the temperature-rise of the disk must stay reasonable to ensure a 1D heat flow and small thermal aberrations. On the other hand, higher doping concentrations also allows us to decrease the thickness of the disk at constant gain, which can lead to better heat removal. In both cases, the limiting factors of increasing the doping concentrations are relevant to evaluate future power scaling and have been little explored in the thin-disk geometry and will most likely be one of the future directions to follow.

Ho:YAG has a much more sophisticated energy level structure (Fig. 8) than the more classically used Yb:YAG, which can lead to various deleterious transitions. This can potentially decrease overall efficiency of the lasing process, depending on doping level. The 1908-nm pump laser excites $Ho^{3+}$ ions from the ground manifold $^5I_8$ to the lasing manifold $^5I_7$. Possible transition processes from the lasing manifold are stimulated emission from $^5I_7$ to $^5I_8$ (main lasing transition), up-conversion and cross-relaxation. Up-conversion is the non-radiative deleterious process caused by interaction between two exited $Ho^{3+}$ ions in the $^5I_7$ manifold, which results in excitation of one of these to the upper levels $^5I_6$ or $^5I_5$ and relaxation of the other to the ground level. This leads to quenching of the upper laser level, reducing the number of ions which can participate in the lasing transition. On the other hand, cross-relaxation is the opposite process, featuring two ions transiting to $^5I_7$ from the $^5I_5$ and $^5I_8$ manifolds, respectively. Considering a low lifetime of $^5I_5$, up-conversion has sufficiently higher probability to occur than cross-relaxation.

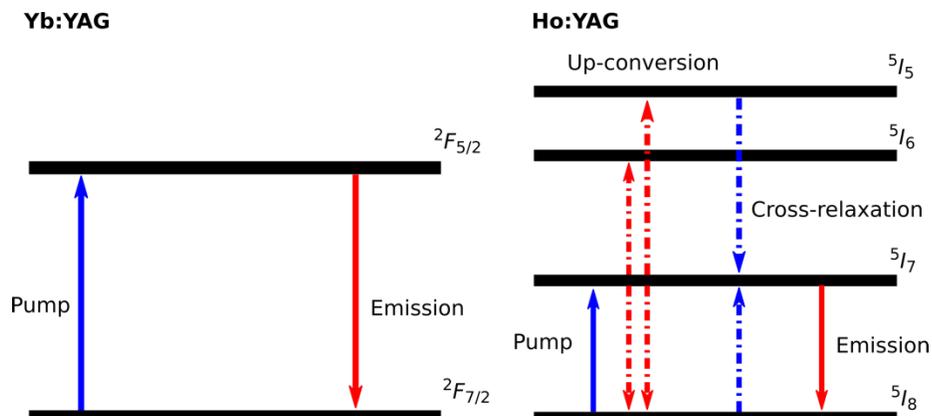

Fig. 8. The comparison of Yb:YAG and Ho:YAG level structure and key transitions

Literature [54,55] shows that the up-conversion probability coefficient increases with increasing the $Ho^{3+}$-doping concentration, which is a major concern on the design of TDLs based on this material. It is also worth noting that up-conversion effects also become more severe at higher temperatures, also emphasizing the importance of the above-mentioned considerations on thickness optimization together with doping concentration.

In [56], it is shown that bulk Ho:YAG crystals with lower doping concentration perform better in terms of threshold and slope efficiency. This can be explained by the presence of two effects – rising up-conversion mentioned above and increasing thermal load due to the higher pump absorption density. Among the other bulk Ho:YAG studies, doping concentrations spread from 0.3 % to 1 % [25,48,57–60], confirming that lower doping concentration are beneficial in this geometry. In contrast to bulk lasers, TDLs allow for significantly higher pumping intensities due to the improved heat removal, but pumping volume is comparatively small. This in turn means that the doping concentration limits in this geometry still need to be experimentally investigated. One important aspect of such an investigation is that, to the best of our knowledge, up-conversion probability values differ widely between different studies presented in literature [54,55] and simulations are only indicative.

All this seems to indicate that disks with higher doping concentration seems to be the next reasonable step for the further power scaling. In [43], a 2.5-at.% doped, 200-µm thick disk led to a higher slope efficiency, but lower output power, showing the potential for further improvements (as illustrated in Fig. 8). In the near future, we plan to perform further experiments with up to 6-at.% doped Ho:YAG thin-disk active media (which are commercially available) and thicknesses varying between 100 µm and 200 µm, and explore in detail the temperature-rise and its influence on laser performance. Higher doping levels can be considered in the future, in combination with thinner disks.

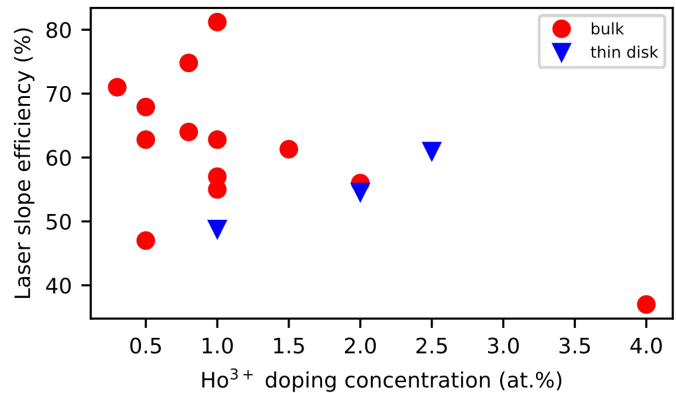

Fig. 9. Laser slope efficiency versus doping concentration for bulk and thin-disk Ho:YAG lasers

3.3 Discussion on the challenges of modelocking at the 100-W level

Our main goal is to develop a 100-W class mode-locked 2-µm TDL, and in this section, we discuss the current challenges.

The first and perhaps the most obvious difficulty is the significantly reduced availability and increased price of commercial optical components and characterization equipment designed for this wavelength range. This, however, is changing very rapidly with the increased interest in high-power laser systems in this wavelength region.

One particularly challenging component are dispersive mirrors with low loss and small thermal effects. This issue has been discussed in the 1-µm regime [50] and is even more severe in the 2-µm region because significantly less knowledge about thermal and damage properties of optics is available. Nevertheless, our experiments show that current state-of-the-art commercially available dispersive mirrors can support 100-W level output. In Fig. 10a, we show a dispersive mirror with -500 fs$^2$ of group delay dispersion (Layertec GmbH) based on a quartz wafer operating as a folding mirror in the cavity with beam radius of 0.57 mm and 1.3 kW of intracavity power, and in Fig. 10b a sapphire-wafer based mirror with the same coating under identical conditions. As expected, the use of sapphire substrates together with optimized structures significantly reduces thermal effects and allow operating at kW-class intracavity powers with strongly reduced thermal effects. This will support our next goal of achieving ultrafast operation.

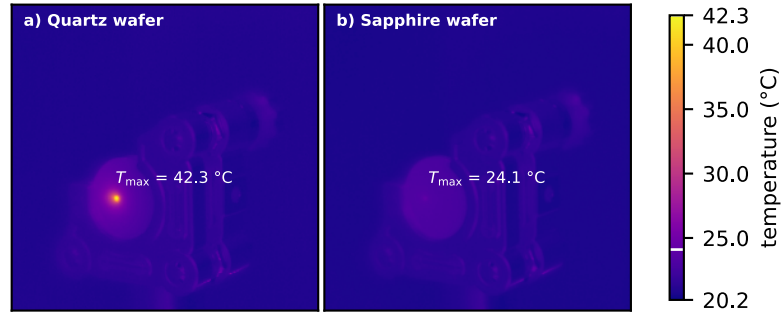

Fig 10. Thermal measurements of dispersive mirrors based on a) quartz and b) sapphire wafers

Other additional difficulties make it more challenging to achieve stable modelocking in Ho:YAG. As we target high output pulse energies, semiconductor saturable absorber mirror (SESAM) modelocking is preferred, however whereas high-power InGaAs-based SESAMs are well established in the 1-µm region [61], GaSb-based SESAMs are only now starting to be explored for higher power operation. We believe however, that promising first steps are currently being achieved in this direction [63]. Another potential issue is that Ho:YAG has a spiked gain profile which can lead to modelocking instabilities when targeting very short pulse durations, imposing additional difficulties in the resonator gain management.

All these limitations make the development of high-power ultrafast 2-µm TDLs challenging. Nevertheless, we believe that we are now at a turning-point in terms of technology development, which will straightforwardly allow to make significant advances in this field. We expect modelocking at the 100-W level to be achievable in the near future.

## 4. Conclusion

In summary, we have conclusively shown the potential of Ho:YAG for high-power operation in the thin-disk geometry. We demonstrated a CW, single fundamental-mode 2-µm Ho:YAG thin-disk oscillator pumped by a commercial 1908 nm Tm-fiber laser, operating at more than 100 W of output power, which is to the best of our knowledge the highest power 2-µm TDL so far. Two different disks with doping concentrations of 1 and 2 % were tested with a set of different OCs. The best performance was achieved with a 2-at.% disk doping concentration and a 3-% transmission OC, reaching 112 W at 53.5-% optical-to-optical efficiency, which corresponds to 3.73 kW of intracavity power, with nearly fundamental beam quality. The slope efficiency in this optimized case was 54.6 %. Caird analysis was performed for the 2-at.% disk based on three different OCs. It resulted in a very low intracavity losses value of 0.8 %, which demonstrates the high quality of the used optical components, proper alignment and absence of strong thermal effects.

In the near future, disks with higher doping concentration, are expected to provide even higher efficiency and gain in the current configuration and allow for better CW performance. In the future, larger pump powers and larger beam sizes on a single-disk are also possible, however limited by the limited availability of cost-efficient Tm-fiber pumps with much higher power levels. Furthermore, we can expect 100-W level modelocked 2-µm lasers to be available in the near future, supported by the above-mentioned higher gain levels, and improvements in components such as dispersive mirrors and SESAMs.

## 5. Acknowledgements

This work was funded by the Deutsche Forschungsgemeinschaft (DFG, German Research Foundation) under Germany's Excellence Strategy – EXC-2033 – Projektnummer 390677874, as well as by European Research Council (ERC) under the European Union's Horizon 2020 research and innovation programme (grant agreement No. 805202 - Project Teraqua). The authors thank Layertec GmbH for fruitful discussions and for designing custom high-power optical coatings for 2-µm wavelength range.